\documentstyle[sprocl]{article}

\def\be{\begin{equation}}
\def\ee{\end{equation}}
\def\bea{\begin{eqnarray}}
\def\eea{\end{eqnarray}}

\begin{document}

\title{TWO-MODE SQUEEZED STATES \\ IN HIGH ENERGY PHYSICS:\\
KEY TO HADRON MODIFICATION IN DENSE MATTER ?
}

\author{M. ASAKAWA}

\address{Department of Physics, School of Science, Nagoya University,\\
Nagoya, 464 - 01, Japan}

\author{\underline{T. CS\"ORG\H O}}

\address{MTA KFKI RMKI, H - 1525 Budapest 114, POB. 49, Hungary \\
	Department of Physics, Columbia University, \\
	538 W 120-th Street, New York, NY 10027, USA}

\maketitle\abstracts{
We discuss the possibility to observe hadron modification in
hot and dense matter via the correlation of identical particles.
We find that a modification of hadronic masses in medium 
leads to two-mode squeezing which signals itself in 
back-to-back correlations of hadrons. Although this effect leads to a
signal of a shift in $\phi$-mass at RHIC, larger signal is expected
for the mass shift of kaons or pions. 
}
  
\section{Introduction}
The Hanbury-Brown Twiss (HBT) effect has been widely measured in heavy ion
collisions. It has been expected that the effect will give some clue to
the size of the system at freeze-out. Another interesting topic in heavy
ion physics is the possibility of hadron modification in medium. So far
these two have been considered as two different aspects of heavy ion physics.
The HBT effect is concerned with freeze-out where interaction disappears,
whereas the hadron modification is caused by interaction. 
However, this picture is purely classical. Since the HBT effect is of quantum
nature, we need quantum mechanical consideration.

In relativistic heavy ion collisions, freeze-out looks rather prompt.
The distribution of final state hadrons is, in most of the cases, almost
exponential~\cite{QM95}
and this suggests that the system is almost thermalized up to
some time and then breaks up suddenly~\cite{cc}. 
Motivated by this, we have modeled~\cite{aw,ac}
freeze-out as follows. The system remains thermalized until freeze-out.
Hadrons are modified due to interaction and their masses are shifted.
As a result, it is dressed pseudo-particles that are thermalized.
Then, freeze-out is assumed to take place suddenly. 
This scenario was investigated for small modification
of pion and kaon masses in ref.~\cite{ac}. Here we summarize the correct
theoretical results, following the ideas of ref.~\cite{ac}, 
and will apply the results to the
correlation signal of mass modification of $\phi$ mesons. 

\section{Calculation of Correlation Functions}
In this paper, we use the following scalar theory and we ignore the isospin
structure, because it is not essential in our argument. Throughout this
paper, we adopt the mean field approximation for simplicity and for clarity.
In addition, we do not take into account the finiteness of the system.
Calculations for finite systems will be presented elsewhere.

The theory in the vacuum is given by the following free Lagrangian
${\cal L}_0$:
\begin{equation}
{\cal L}_0 = \frac{1}{2}\partial_{\mu}\phi(x)\partial^{\mu}\phi(x)
-\frac{1}{2}m_0^2 \phi^2(x),
\end{equation}
where $m_0$ is the vacuum mass of the scalar field $\phi(x)$.
After standard canonical quantization and 
normal ordering, we get the well-known diagonalized Hamiltonian,
$
H_0=\int\! \omega_{\bf k}^{\phantom{\dagger}}a^{\dagger}_{\bf k}a_{\bf k}^{\phantom{\dagger}} d^3 {\bf k},
$
where $\omega_{\bf k}^{\phantom{\dagger}} = \sqrt{m_0^2 + {\bf k}^2}$. 
When the temperature and/or chemical density is non-vanishing and
the mass of the $\phi$ field is shifted, within the mean field
approximation, this is expressed by the following Lagrangian in
medium ${\cal L}_M$:
\begin{equation}\label{lmed}
{\cal L}_M = \frac{1}{2}\partial_{\mu}\phi(x)\partial^{\mu}\phi(x)
-\frac{1}{2} m_*^2 \phi^2(x).
\end{equation}
The mass shift $\delta M$ is given by
$
\delta M = m_*  - m_0 .
$
Let us define the quanta that diagonalize the Lagrangian (\ref{lmed})
as $b_{\bf k}^{\phantom{\dagger}}$. The point here is that the $b$-quanta are, in general,
different from the $a$-quanta which are the fundamental excitations
in the vacuum. In other words, the $a$-operators do not diagonalize
the Hamiltonian in medium $H_M$. By normal ordering of $H_M$,
 we obtain
\begin{eqnarray}
H_M & = & H_0 + H_1, \nonumber \\
H_1 & = & \frac{m_*^2 - m_0^2}{4}\int \!\frac{1}
	{\omega_{\bf k}}
[ a_{\bf k}^{\phantom{\dagger}}
  a_{\bf -k}^{\phantom{\dagger}}
+ 2a^{\dagger}_{\bf -k}a_{\bf -k}^{\phantom{\dagger}}
	+a^{\dagger}_{\bf k} a^{\dagger}_{\bf -k} ]
\, d^3 {\bf k}. \label{hmed}
\end{eqnarray}
Therefore, in medium, mode ${\bf k}$ and mode ${\bf -k}$ of the $a$-quanta
are mixed. This Hamiltonian (\ref{hmed}) can be exactly diagonalized with
the following Bogoliubov transformation:
\begin{eqnarray}
a^{\dagger}_{\bf k}
	\!\!\!\!&=&\!\!\!\!
\cosh r_{\bf k}^{\phantom{\dagger}}\, b^{\dagger}_{\bf k}
+\sinh r_{\bf k}^{\phantom{\dagger}}\, b_{\bf -k}^{\phantom{\dagger}}, \,\,
\quad 
a^{\dagger}_{\bf -k} = \cosh r_{\bf -k}^{\phantom{\dagger}}\,
b^{\dagger}_{\bf -k} +\sinh r_{\bf -k}^{\phantom{\dagger}}\, b_{\bf k}^{\phantom{\dagger}},
\nonumber \\ 
a_{\bf k}^{\phantom{\dagger}}\!\!\!\!& = & \!\!\!\!\sinh r_{\bf k}^{\phantom{\dagger}}\,
b^{\dagger}_{\bf -k} +\cosh r_{\bf k}^{\phantom{\dagger}}\, b_{\bf k}^{\phantom{\dagger}}, \,\,
\quad
a_{\bf -k} = \sinh r_{\bf -k}^{\phantom{\dagger}}\,
b^{\dagger}_{\bf k}
+\cosh r_{\bf -k}^{\phantom{\dagger}} \, b_{\bf -k}^{\phantom{\dagger}},  
\label{bog}
\end{eqnarray}
where 
$r_{\bf k}^{\phantom{\dagger}} = 0.5 \log\left(\omega_{\bf k}^{\phantom{\dagger}}/\Omega_{\bf k}^{\phantom{\dagger}} \right)$
and 
$
\Omega_{\bf k}^{\phantom{\dagger}} = \sqrt{m_*^2 +  {\bf k}^2}.
$
This gives the exact relationship between the quanta in the vacuum
($a$-quanta) and the quanta in medium ($b$-quanta) in this theory.
With the $b$-operators, the in-medium Hamiltonian (\ref{hmed}) is
diagonalized as
$
H_M=\int\! \Omega_{\bf k}^{\phantom{\dagger}}b^{\dagger}_{\bf k}
	b_{\bf k}^{\phantom{\dagger}} d^3 {\bf k}.
$
This is almost trivial, since the mass of the $\phi$ field is
shifted to $m_*$ in medium. However, the important point
here is that it is the $b$-quanta that are thermalized in medium, since the
$b$-operators diagonalize the in-medium Hamiltonian. On the other hand,
the $b$-quanta are not observed experimentally. It is the $a$-quanta
that are observed. Therefore, in calculating final state observables,
we have to evaluate the expectation value of operators defined
in terms of $a$ and $a^{\dagger}$ operators,
${\cal O}(a, a^{\dagger})$
with the density matrix defined in the $b$-basis, $\rho_b$, i.e.,
$
\langle {\cal O}(a, a^{\dagger})\rangle
= {\rm Tr}\,\rho_b {\cal O}(a, a^{\dagger}).
$
The calculation of the one and two-particle distribution functions
which are needed to obtain the two-particle correlation is straightforward,
but the Glauber -- Sudar\-shan representation of the thermal density matrix
\cite{gs},
\begin{equation}
\rho_b = \prod_{\bf k}^{\phantom{\dagger}} \int\! \frac{d^2\beta_{\bf k}^{\phantom{\dagger}}}{\pi}
P(\beta_{\bf k}^{\phantom{\dagger}})
|\beta_{\bf k}^{\phantom{\dagger}}\rangle\langle\beta_{\bf k}^{\phantom{\dagger}}|,
\end{equation}
is useful \cite{aw}.
 Here $|\beta_{\bf k}^{\phantom{\dagger}}\rangle$ is a coherent state satisfying
$b_{\bf k}^{\phantom{\dagger}} |\beta_{\bf k}^{\phantom{\dagger}}\rangle = \beta_{\bf k}^{\phantom{\dagger}} |\beta_{\bf k}^{\phantom{\dagger}}\rangle$
and 
\begin{equation}
P(\beta_{\bf k}^{\phantom{\dagger}}) = 
	\frac{1}{n_{\bf k}}
\exp\left (-\frac{|\beta_{\bf k}^{\phantom{\dagger}}|^2}
	{n_{\bf k} }\right ),
\end{equation}
where $n_{\bf k}^{\phantom{\dagger}}$ is a Bose--Einstein distribution function with
mass $m_*$:
\begin{equation}
n_{\bf k} = {1 \over \exp(\sqrt{m_*^2 + {\bf k}^2 }/T ) - 1 }.
\end{equation}
We obtain the following measurable one-particle 
distribution in the final state:
\begin{equation}
N_1({\bf k}) \, = 
	\, \langle a^{\dagger}_{\bf k} a_{\bf k}^{\phantom{\dagger}} \rangle
= | c_{\bf k}^{\phantom{\dagger}}|^2 \, n_{\bf k}^{\phantom{\dagger}} + 
		|  s_{- \bf k}|^2 n_{- \bf k} + 
		| s_{-\bf k}|^2 ,
	\label{e:spect}
\end{equation}
where $c_{\bf k}^{\phantom{\dagger}} = \cosh r_{\bf k}^{\phantom{\dagger}}$, $s_{\bf k}^{\phantom{\dagger}} = \sinh r_{\bf k}^{\phantom{\dagger}}$ (similar
expressions hold for
${- \bf k}$).
If we assume the uniformity of the system, i.e.,
$n_{\bf k}^{\phantom{\dagger}}=n_{-{\bf k}}^{\phantom{\dagger}}$ and 
$r_{\bf k}^{\phantom{\dagger}}=r_{\bf -k}^{\phantom{\dagger}}$,
the spectrum of eq.~(\ref{e:spect}) will become 
the same as  the spectrum obtained for one-mode squeezed states~\cite{ac}.
Since we have ignored the finite size effect, 
the two-particle distribution
function
$\langle a^{\dagger}_{\bf k} a^{\dagger}_{{\bf k}'}
a_{\bf k}^{\phantom{\dagger}}a_{{\bf k}'}^{\phantom{\dagger}} \rangle $ takes a trivial value
$\langle a^{\dagger}_{\bf k}
	a_{\bf k}^{\phantom{\dagger}} \rangle 
\langle a^{\dagger}_{{\bf k}'}a_{{\bf k}'}^{\phantom{\dagger}} \rangle$
unless ${\bf k}=\pm{\bf k}'$.
We get the non-trivial two-particle correlation functions
$C_2({\bf k}, \pm{\bf k})$ as follows:
\begin{eqnarray}
C_2({\bf k}, {\bf k}) & = & 2 (!), \nonumber \label{e:c2}\\
C_2({\bf k}, {\bf -k}) & = & 1 +
{\displaystyle\phantom{|} {|c^*_{\bf k}s_{\bf k}^{\phantom{\dagger}}
	n_{\bf k}^{\phantom{\dagger}}+
	c^*_{-\bf k} s_{-\bf k}^{\phantom{\dagger}} n_{-\bf k}^{\phantom{\dagger}} +
	c^*_{-\bf k} s_{-\bf k}^{\phantom{\dagger}}|^2 }
\over \displaystyle\phantom{|} {N_1({\bf k}) \, N_1({- \bf k})} }
\, \neq \, 1 \ . 
	\label{e:c2b}
\end{eqnarray}
In all other cases, the two-particle correlation function is 1.
This result implies two important issues,
(i) the intercept of the correlation function
$C_2({\bf k}, {\bf k})$ remains the canonical value of 2,
even if quanta in medium are different from those in the vacuum,
 (ii) back-to-back
correlation 
($C_2({\bf k}, {\bf -k}) \neq 1$)
is generated by hadron modification. This is caused by the mixing of
${\bf k}$ and ${\bf -k}$ modes due to the mean field effect. 

\begin{center}
\vspace*{7.5cm}
\includegraphics{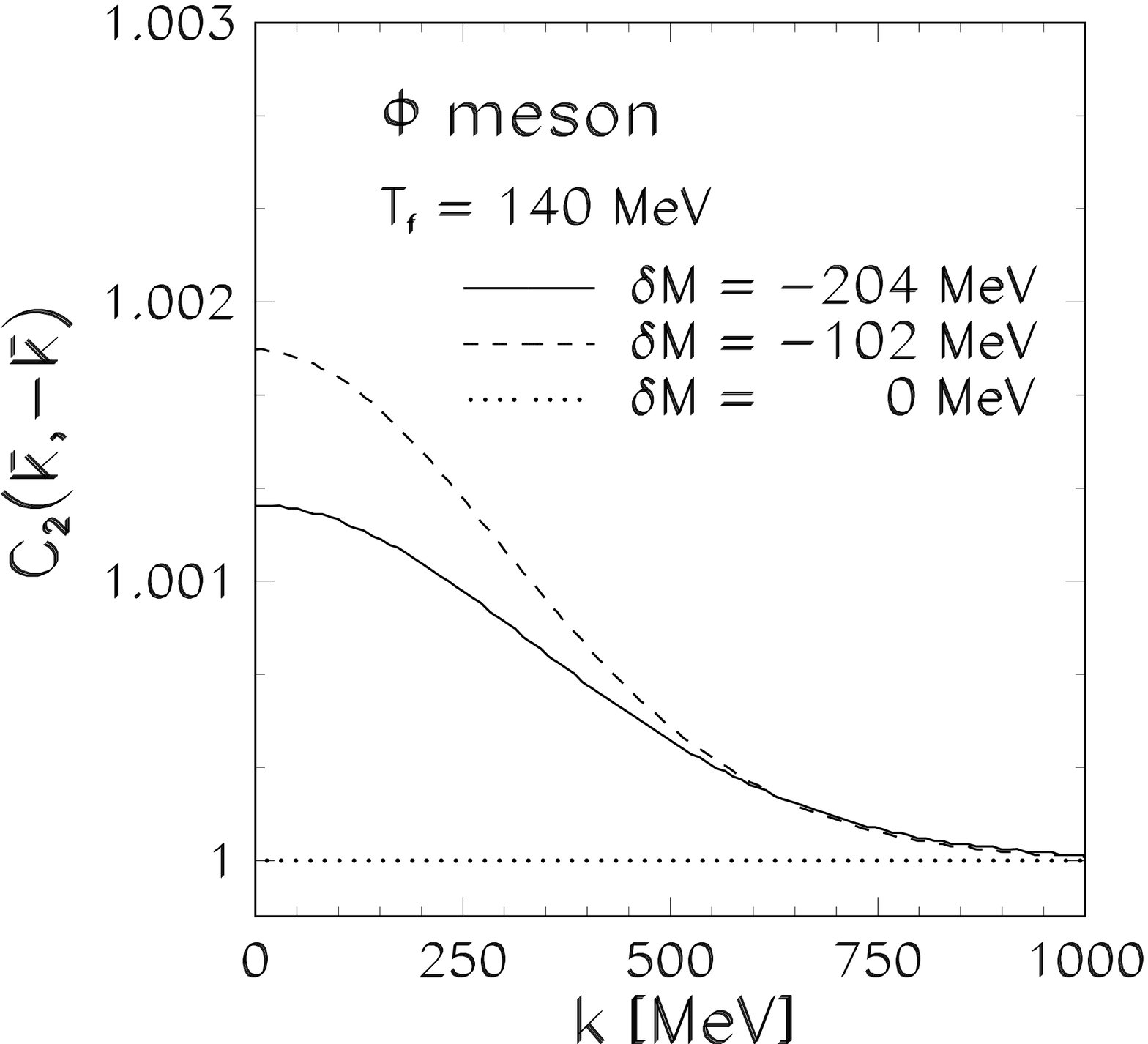}
\begin{minipage}[t]{10.054cm}
{\small {\bf Fig.~1.}
Large mass-shift of $\phi$ results in a small back-to-back correlation.
Solid line stands for a 10 \%, dashed line for a
20 \% decrease in $m_{\phi}$. 
The freeze-out temperature was assumed to be $ T = 140 $ MeV.
Similar effect for pions leads to bigger signal.}
\end{minipage}
\end{center}

Thus the correlation properties of two-mode squeezed states are essentially
different from those properties of the one-mode squeezed states, 
that were invoked in ref.~\cite{aw}. 
In our case,
the mean field carries no momentum. Due to momentum conservation, 
only ${\bf k}$ and ${\bf -k}$ modes get additional correlation.
In contrast, when one mode squeezing is assumed, 
 $C_2({\bf k}, {\bf -k}) = 1$ and one can show that the intercept of the
two-particle Bose-Einstein correlation function may take up
any non-negative value: $ 0 < C_2({\bf k},{\bf k}) < \infty$.
Although the importance of the back-to-back correlation functions
was well realized in ref~\cite{ac}, the formula for the back-to-back
correlation function was incorrectly given by eq. (18) of ref.~\cite{ac}.  
The correct  expression for the 
back-to-back correlation function 
is presented by our eq.~(\ref{e:c2b}).

Since back-to-back correlations of $\phi$ mesons may be observable
by the PHENIX detector at RHIC~\cite{lz},
(the Relativistic Heavy Ion Collider which shall be operational
by 1999 colliding $Au + Au$ at $\sqrt{s} = 40 $ TeV) we apply here
our model to the case of a 10 - 20 \% mass-shift of $\phi$-mesons on Fig. 1.
The back-to-back correlation function  of $\phi$ mesons has a  worse
signal-to-noise ratio, than that of kaons, which is due to 
the heavier mass of $\phi$ ($m_{\phi} = 1020 $ MeV vs. $m_K = 494$ MeV). 
Fig. 1. indicates that a 10 - 20 \% shift in the
mass of $\phi$ mesons results in a small  signal in the back-to-back
correlation function.

\section{Summary}

 A new method has been found to test the medium modification of
bosons, utilizing their quantum correlations at large
momentum difference. The effect follows from basic principles
of statistical physics and canonical quantization.

 Back-to-back correlations are not contaminated by resonance decays at larger
momentuma of the particles, in
contrast to the Bose--Einstein correlation functions at small
relative momenta.
For a locally thermalized expanding source, however,  
back-to-back correlations will appear in the rest frame 
of each fluid element. Thus, for such  
systems a more realistic estimate is necessary to learn more
about the magnitude of the correlations of $\phi$ or other mesons 
at large relative momenta. 
Contrary to previous expectations~\cite{aw,ac} the effect is the biggest
for the lightest particles (photons, pions or kaons)  and the back-to-back
correlations vanish for very large values of particle momenta. 
In the present study we have 
neglected finite size effects, which will make the 
correlation function vary smoothly around both ${\bf k_1} = {\bf k_2}$
and ${\bf k_1} =  - {\bf  k_2}$.

\vskip 15pt

%
\section*{Acknowledgment}
\noindent
We would like to thank M. Gyulassy and J. Knoll for enlightening
discussions.  \mbox{T. Cs\"org\H o} is grateful to  Gy\"orgyi and M. Gyulassy for
kind hospitality at Columbia University.
This research has been supported by the US - Hungarian Joint Fund
under contract MAKA 378/1993, by the Hungarian NSF under contract
OTKA - F 4019, T024094 and W01015107 and by an 
Advanced Research Award of the  Fulbright Foundation, grant document 
20926/1996.

\vfill\eject 
\end{document}